\documentclass[
prl,
 reprint,
 amsmath,amssymb,
 aps,preprintnumbers]{revtex4-1}

\usepackage[colorlinks=true,linkcolor=black, citecolor=black,
urlcolor=black]{hyperref}

\usepackage{multirow,graphics}
    \newcommand{\beq}{\begin{equation}}
    \newcommand{\eeq}{\end{equation}}
    \newcommand\beqa{\begin{eqnarray}}
    \newcommand\eeqa{\end{eqnarray}}

\usepackage{amstext}
\usepackage{amssymb}
\usepackage{amsmath}
\usepackage{graphicx}
\usepackage{color}

\begin{document}

\newcommand{\IM}{{\rm Im}\,}
\newcommand{\card}{\#}
\newcommand{\la}[1]{\label{#1}}
\newcommand{\eq}[1]{(\ref{#1})}
\newcommand{\figref}[1]{Fig \ref{#1}}
\newcommand{\abs}[1]{\left|#1\right|}
\newcommand{\comD}[1]{{\color{red}#1\color{black}}}

\makeatletter
     \@ifundefined{usebibtex}{\newcommand{\ifbibtexelse}[2]{#2}} {\newcommand{\ifbibtexelse}[2]{#1}}
\makeatother

\newcommand{\footnoteab}[2]{\ifbibtexelse{
\footnotetext{#1}
\footnotetext{#2}
\cite{Note1,Note2}
}{
\newcommand{\textfootnotea}{#1}
\newcommand{\textfootnoteab}{#2}
\cite{thefootnotea,thefootnoteab}}}
\newcommand{\footnoteb}[1]{\ifbibtexelse{\footnote{#1}}{
\newcommand{\textfootnoteb}{#1}
\cite{thefootnoteb}}}
\newcommand{\footnotebis}{\ifbibtexelse{\footnotemark[\value{footnote}]}{
\cite{thefootnoteb}}}

\def\bnu{{\bar{\nu}}}
\def\umu{{\underline{\mu}}}
\def\e{\epsilon}
     \def\bT{{\bf T}}
    \def\bQ{{\bf Q}}
    \def\wT{{\mathbb{T}}}
    \def\wQ{{\mathbb{Q}}}
    \def\ttQ{{\bar Q}}
    \def\tQ{{\tilde \bP}}
        \def\bP{{\bf P}}
    \def\CF{{\cal F}}
    \def\cC{\CF}
     \def\Tr{\text{Tr}}
     \def\l{\lambda}
\def\hbZ{{\widehat{ Z}}}
\def\bZ{{\resizebox{0.28cm}{0.33cm}{$\hspace{0.03cm}\check {\hspace{-0.03cm}\resizebox{0.14cm}{0.18cm}{$Z$}}$}}}
\newcommand{\rb}{\right)}
\newcommand{\lb}{\left(}
\newcommand{\nn}{\nonumber}

\newcommand{\gT}{T}\newcommand{\gQ}{Q}

\title{ The Quantum Spectral Curve of the ABJM Theory }

\author{ Andrea Cavagli\`a$^{a}$ , Davide Fioravanti$^{b}$ , Nikolay Gromov$^{c}$ and Roberto Tateo$^{a}$}

\affiliation{
\(^{a}\) Dip.\ di Fisica
and INFN, Universit\`a di Torino, Via P.\ Giuria 1, 10125 Torino, Italy
\\
             \(^{b}\) INFN-Bologna and Dipartimento di Fisica e Astronomia,
Universit\`a di Bologna, Via Irnerio 46, 40126 Bologna, Italy
\\
\(^{c}\)Mathematics Department, King's College London, The Strand, London WC2R 2LS, UK and
 St.Petersburg INP, Gatchina, 188300, St.Petersburg, Russia
}
\begin{abstract}
Recently, it was shown that the spectrum of anomalous dimensions and other important observables in $\mathcal{N}=4$ SYM are encoded into a simple nonlinear Riemann-Hilbert problem: the $\bf P\mu$-system or Quantum Spectral Curve. In this letter we present the $\bf P\mu$-system for the spectrum of the ABJM theory. This
may be an important step towards the exact determination of the interpolating function $h(\lambda)$ characterising the integrability of the ABJM model. We also discuss a surprising symmetry between the $\bP \mu$-system equations for $\mathcal{N}=4$ SYM and ABJM.
\end{abstract}

 \maketitle

\section{Introduction}
\begin{figure}
  \centering
  \includegraphics[scale=0.3]{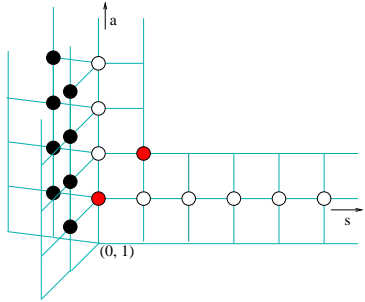}
  \caption{T-hook for the ABJM T-system }
 \label{fig:THook}
\end{figure}
The ABJM model~\cite{ABJM} is a unique example of three dimensional gauge theory which may be completely solvable in the planar limit. In particular, echoing the developments in the study of $\mathcal{N}=4$ Super Yang-Mills theory in 4d, an exact description of the spectrum of conformal dimensions has been obtained by combining information from two-loop perturbation theory \cite{MinahanZarembo} and on the strong coupling limit, corresponding to the classical limit of type
IIA superstring theory
on $AdS_4 \times CP^3$~\cite{Stefanski:2008ik,Arutyunov:2008if,ClassicalCurve}. This lead to the Asymptotic Bethe Ansatz conjectured in \cite{ABA}, describing operators with large quantum numbers, and ultimately to the Thermodynamic Bethe Ansatz (TBA) equations~\cite{TBA1,TBA2}, an infinite set of nonlinear integral equations encoding the  anomalous dimensions spectrum as a function of a dressed coupling constant $h(\lambda)$. Finding the exact dependence of $h$ on the t'Hooft coupling $\lambda$ is still a missing link in the integrability approach to the ABJM theory (see \cite{Klose:2010ki} for a review).

It is expected that other important observables can be studied with integrable model tools. In the case of $\mathcal{N}=4$ SYM, it was shown in
\cite{MaldacenaCusp,Drukker:2012de} that a system of Boundary Thermodynamic Bethe Ansatz equations describes the (generalised) cusp anomalous dimension $\Gamma(\phi)$ characterising the logarithmic UV divergences of light-like Wilson lines forming a cusp of angle $\phi$. In some near-BPS limits, the cusp anomalous dimension can also be studied with independent localisation techniques (see for example \cite{Correa:2012at}), leading to non-perturbative exact results which nicely agree with integrability
 computations~\cite{Gromov:2012eu,Gromov:2013qga}.

 For the ABJM model, the Bremsstrahlung function $B(\lambda)$ characterising the leading small angle behaviour $\Gamma(\phi) \sim \phi^2 B(\lambda)$ was recently computed
 in~\cite{LastMaldacena} (see also \cite{BPSWilsonLoops} for related results).
 As already put forward in \cite{MaldacenaCusp}, obtaining the same quantity with integrability methods would allow to fix the exact relation between $h$ and $\lambda$.

An important development in $\mathcal{N}=4$ SYM was the discovery of an alternative formulation of the TBA as
a nonlinear matrix Riemann-Hilbert problem, known as $\bf P \mu$-system or Quantum Spectral Curve (QSC). It is a finite set of universal functional relations, believed to encode not only all states of the anomalous dimension spectrum, but also, with an appropriate change in the asymptotics, the cusp
spectrum~\cite{PmuAdS5,Gromov:2013qga}.
This new tool
also proved to be much more efficient than the TBA for extracting exact results. In particular it lead to the 9 loop prediction for the Konishi dimension at weak
coupling~\cite{9loops},
 $3$ loops at strong coupling~\cite{QSCAtWork} as well as to new results in the study of BFKL pomeron.

In this letter we present the ${ \bf P \mu}$-system for the ABJM theory, and discuss a surprising link with the Quantum Spectral Curve equations for $\mathcal{N}=4$ SYM.

While here we only discuss the application of this new set of equations to the spectrum of anomalous dimensions, we believe that it will play an important r$\hat{\text{o}}$le in fixing the $h-\lambda$ relation.

\begin{figure}
\begin{minipage}{0.2\textwidth}
 \centering
  \includegraphics[scale=0.24]{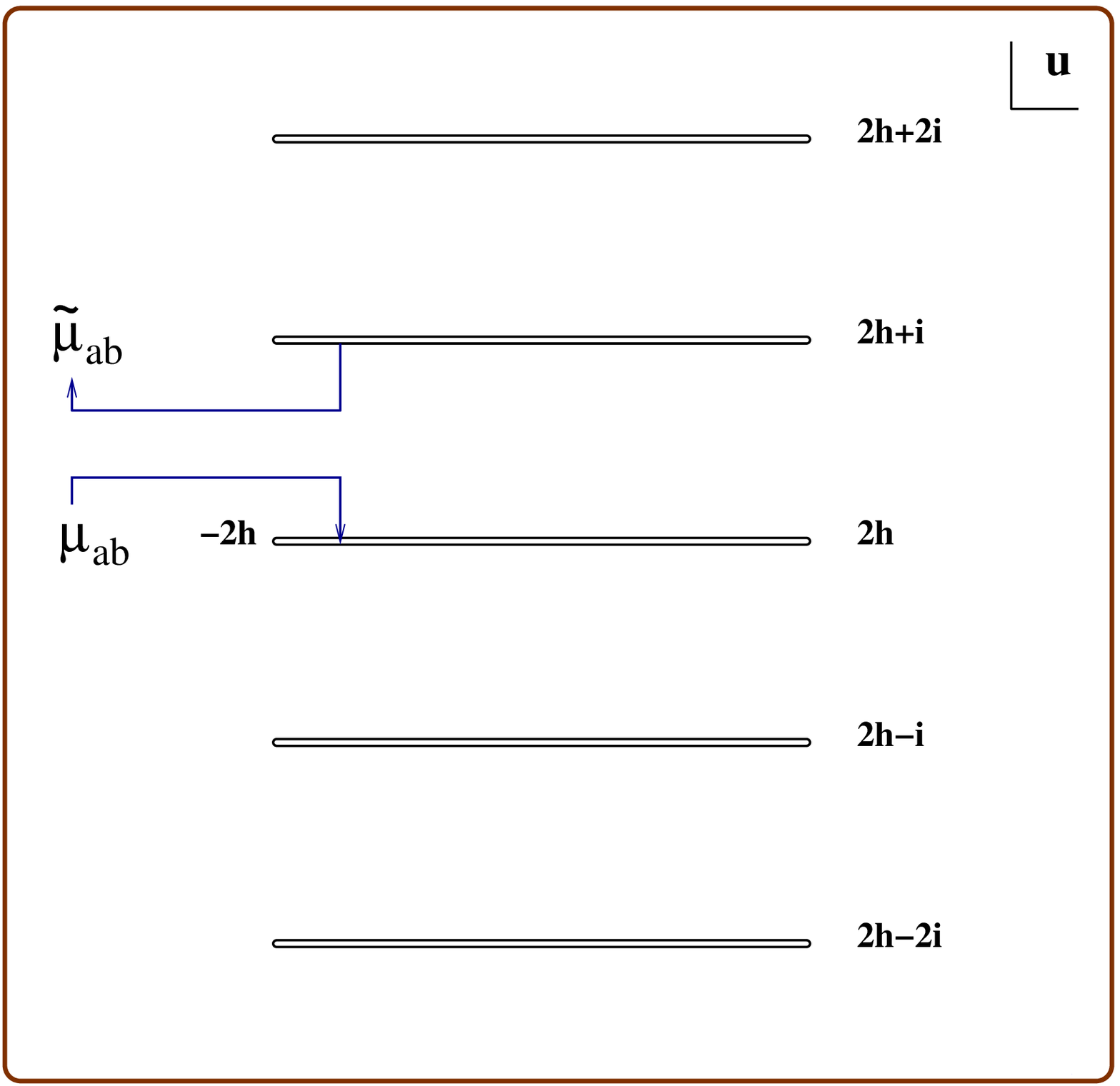}
\end{minipage}
\hspace{0.5 cm}
\begin{minipage}{0.2 \textwidth}
 \centering
  \includegraphics[scale=0.24]{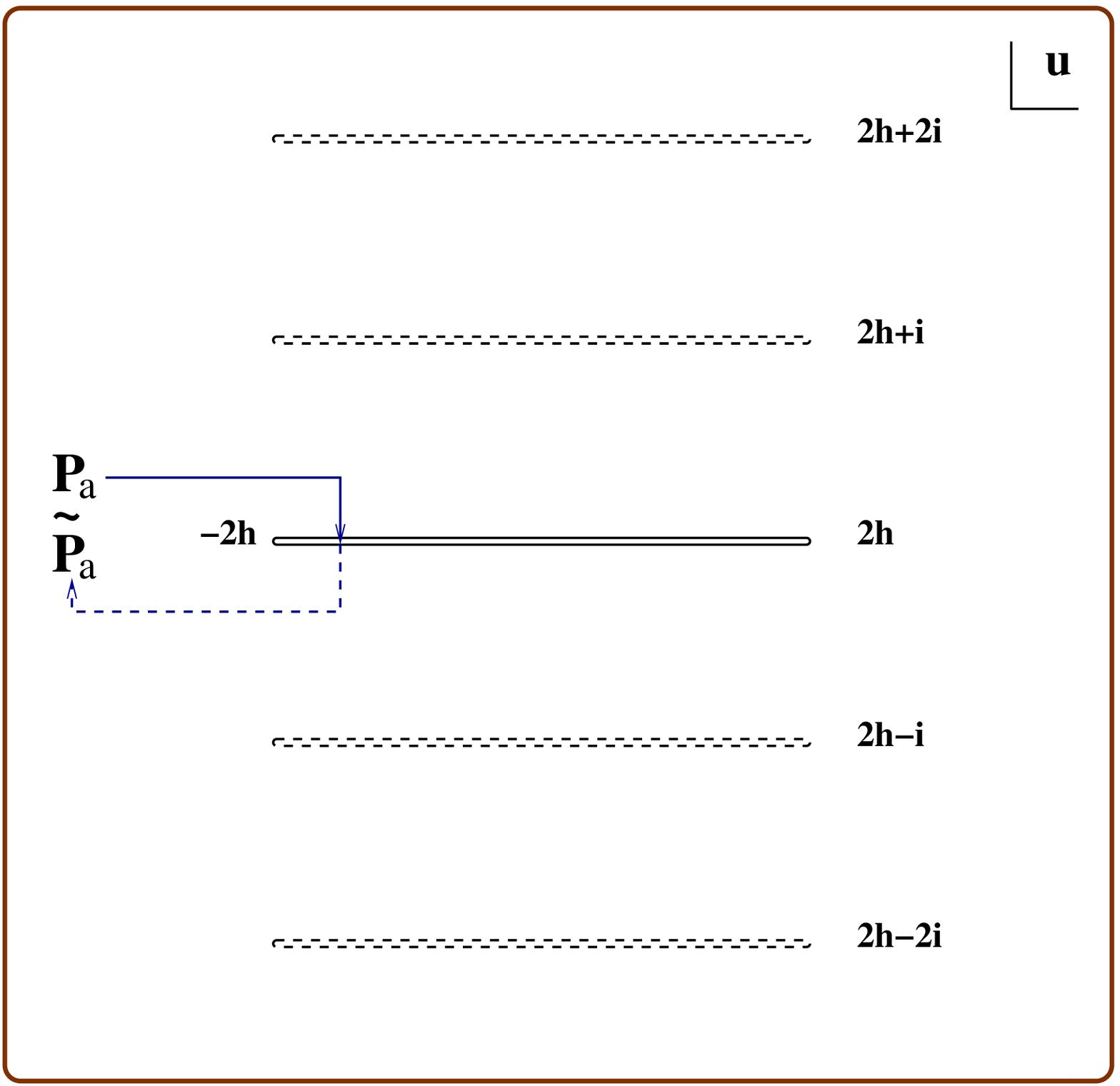}
\end{minipage}
\caption{ Analytic structure for the two types of variables in the Quantum Spectral Curve. }
\label{fig:cuts}
\end{figure}

\section{Outline of the derivation}

Conceptually, the $\bf P \mu$-system is equivalent to other reformulations of the TBA as a set of functional relations, such as the Y- or T-system. In particular it can be derived from the
Y-system~\cite{GKV} supplemented by the discontinuity equations~\cite{Extended,ABJMDisc} describing the monodromies of the Y functions around infinitely many branch points in the complex plane of the spectral parameter $u$. These branch points are located at rigid positions $u = \pm 2 h + i n/2$, $n \in \mathbb{Z}$.
However these relations are very intricate, while the $\bf P \mu$-system involves only a finite number of objects, with the transparent analytic properties shown in Figure
\ref{fig:cuts}~\cite{PmuAdS5}: the ${\bf P }_a$ functions are defined on a Riemann sheet with a single cut running from $-2 h $ to $+2 h$, while the functions $\mu_{ab}$, although still having an infinity of branch cuts for $(-2 h , +2h ) + i n$, $n  \in \mathbb{Z}$, satisfy the simple relation
\beqa\label{eq:mu}
{ \tilde \mu }_{ab}(u) = \mu_{ab}( u + i ) ,
\eeqa
where $\tilde{\mu}$ and ${ \tilde \bP}$ denote the values of the $\bP \mu$ variables analytically continued around one of the branch points on the real axis. Equation (\ref{eq:mu}) means that, on a different Riemann section,
$\mu_{ab}$ is simply an $i$-periodic function~\cite{PmuAdS5}.\\
To reveal this hidden structure, one can start from the analytic properties of the T functions. The T-system for the ABJM spectral problem is defined on the T-hook diagram of Figure
\ref{fig:THook}~\cite{GKV}, where to every node is associated a $T_{a, s}$ function. The latter satisfy the discrete Hirota equation
\beqa\label{eq:TT}
T^{[+1]}_{a, s} T^{[-1]}_{a, s} = \prod_{(a' \sim a )_{\updownarrow} } T_{a', s} + \prod_{(s' \sim s )_{\leftrightarrow} } T_{a, s'},
\eeqa
where the products are over horizontal ({\scriptsize$\leftrightarrow$})  and vertical ({\scriptsize$\updownarrow$}) neighbouring nodes
and $T^{[n]}:= T( u + \frac{i}{2}n )$.

In \cite{SolvingYsystem}, it was discovered a beautiful fundamental set of analyticity conditions for the T functions, and this was adapted to the ABJM case
in \cite{ABJMDisc}, see Appendix C of that paper.
Exploiting the gauge invariance of Hirota equation, it is possible to introduce two very special gauges, denoted as $\bT$ and $\wT$.  For $s \geq a $, the $\wT_{a,s}$ functions can be parametrised as
\beqa\label{eq:whitegauge}
\mathbb{T}_{1 , s} &=& \bP_1^{ \left[ +s \right] } \bP_2^{ \left[ -s \right] } - \bP_2^{ \left[ +s \right] } \bP_1^{ \left[ -s \right] } , \;\;\;\;\; \mathbb{T}_{0 , s} = 1 , \nn\\
\mathbb{T}_{2 , s} &=& \mathbb{T}_{ 1 , 1 }^{ \left[ +s \right] }  \mathbb{T}_{1 , 1 }^{ \left[ -s \right]},
 \;\;\;\;\;\mathbb{T}_{3, 2} / \mathbb{T}_{2,3} = \mu_{12},
\eeqa
where ${\bf P}_1$, ${\bf P}_2$ and $\mu_{12}$ have the simple properties discussed above and will be part of the $\bP\mu$-system.
Furthermore, the $\bf T$ gauge can be introduced with the transformation:
\beqa\label{eq:bbtrans}
{\bf T }_{n , s }  &=& (-1)^{n(s+1)} \mathbb{T}_{n , s}
\left(  {\mu_{12} }^{\left[ n + s - 1 \right] } \right)^{ 2 - n  } ,\hspace{1cm}s\geq 1\nn\\
{\bf T }_{n , 0 }^{\alpha} &=& (-1)^{n}
\mathbb{T}_{n , 0}^{\alpha} \left( \sqrt{ { \mu_{12} }^{\left[ n - 1\right] } } \right)^{ 2 - n  } ,\nn\\
{\bf T }_{n , -1 }^{\alpha} &=& \mathbb{T}_{n , -1}^{\alpha}  = 1, \hspace{0.5cm}{  \scriptsize \alpha = I, II },
\eeqa
and
the ${\bf T }_{n, s}$ functions are required to satisfy
\beqa\label{eq:blackgauge}
{\bf T}_{n, 0}^{\alpha} & \in & \mathcal{A}_{n+1} , \hspace{0.5cm}{ \scriptsize \alpha = I, II } ,\;\;\; n \geq 0 \nn\\
{\bf T}_{n, 1} & \in & \mathcal{A}_{n} , \;\;\; n \geq 1 ,
\eeqa
where we denote with $\mathcal{A}_n$ the class of functions free of branch cuts in the strip $|\text{Im}(u) | < \frac{n}{2} $.

The strategy to derive the $\bf P \mu$-system, to be described in detail in \cite{AdS5Long} and \cite{ABJMLong}, is then the following:
starting from Hirota equation and the gauge transformation (\ref{eq:bbtrans}), it is possible to compute any ${\bf T}_{n, s}$ function in terms of the only variables $\bP_1$, $\bP_2$, $\mu_{12}$, evaluated on different Riemann sheets.
Surprisingly, when rewritten in terms of these functions, the conditions (\ref{eq:blackgauge}) show precisely how the system can be closed introducing only a finite number of fundamental variables, each with one of the two types of cut structures shown in Figure \ref{fig:cuts}.\\

The simplest nontrivial example is provided by the condition
${\bf T}_{2, 1} \in \mathcal{A}_2$. Computing ${ \bf T}_{2, 1}$ as described above, and imposing that it has no cut on the real axis, we find the constraint
\beqa
0 &=& { \bf T}_{2, 1} - { \tilde{ \bf T } }_{2, 1} \\
&=& ( \bP_1^{ \left[ + 2 \right] }
\bP_2^{ \left[ -2 \right] } - \bP_2^{ \left[ + 2 \right] } \bP_1^{ \left[ -2 \right] } )\left( { \tilde \mu }_{12} -
\mu_{12} - \bP_1 {\tilde \bP}_2 +   \bP_2 {\tilde \bP}_1  \right) \nn
\eeqa
The first factor equals $\mathbb{T}_{1, 2}$, which is different from zero, and this leads to a new relation:
\beqa
{ \tilde \mu }_{12} - \mu_{12} = \bP_1 { \tilde \bP }_2 - \bP_2 { \tilde \bP }_1 .
\eeqa
As will be shown in detail in \cite{ABJMLong} (see \cite{AdS5Long} for $\mathcal{N}=4$ SYM), the structure of the $ \bP \mu$-system is already revealed just by inspecting a few of the other conditions in (\ref{eq:blackgauge}).

\section{ The $ \bP\mu$-system }
The ${\bf P }\mu$-system for the ABJM model involves a vector of six functions ${\bf P }_i$, $i = 1, \dots, 6$ and an anti-symmetric $6 \times 6$ matrix $\mu_{ab}$, with the analytic properties of Figure \ref{fig:cuts}.
These variables moreover satisfy the nonlinear constraints
\beqa\la{nonsymConstr}
\bP_5 \bP_6 &=& 1 + \bP_2 \bP_3 - \bP_1 \bP_4, \label{eq:constr1}\\
\mu \chi \mu \chi &=& 0. \label{eq:constr2}
\eeqa
The fundamental Riemann-Hilbert relations contain a $6 \times 6$ symmetric matrix $\chi$ whose only nonzero entries are
\beqa\label{eq:chi}
\chi^{14} = \chi^{41} = -1 , \;\;\; \chi^{23} = \chi^{32} =  1 , \;\;\; \chi^{56} = \chi^{65} = -1 , \nn
\eeqa
and read:
\beqa
\tilde { \bP }_a &=& { \bP }_a - \mu_{ab}\chi^{bc} { \bf P }_c , \label{eq:Pmu1}\\
\mu_{ab}-\tilde \mu_{ab}&=&- { \bf P }_a\tilde { \bf P }_b + { \bf P }_b\tilde { \bf P }_a . \label{eq:Pmu2}
\eeqa
By appropriately tuning the asymptotics, all states of the spectrum can be described by equations (\ref{eq:constr1}-\ref{eq:Pmu2}). 
Below we will discuss the asymptotics only for a specific subsector, postponing the 
general case to a future work~\cite{ABJMLong}.
\paragraph{ Even-parity states. }
For many applications it is useful to consider a reduced system of equations.
As will be shown in \cite{ABJMLong}, the symmetric, parity invariant sector of the spectrum is identified
by the conditions $ \bP_5 = \bP_6$, and $\mu_{5 a } = \mu_{6 a } = -\mu_{a5} = -\mu_{a6}$.
\paragraph{ ${\bf Q }\omega$-system. }
Finally, we remark that, similar to the $\mathcal{N}=4$ case, there is a complementary set of conditions, named $\bf Q \omega$-system~\cite{ABJMLong}, which is formally the same as (\ref{eq:constr1}-\ref{eq:Pmu2}) with the replacements
\beqa
\bP_a \rightarrow {\bf Q}_a , \hspace{1cm} \mu_{ab} \rightarrow\omega_{ab} ,
\eeqa
but with all the branch cuts reversed. Namely, the functions ${ \bf Q}_a$ have a single branch cut for $u \in (-\infty, -2 \; h ) \cup (+2 \; h , + \infty )$, while $\omega_{ab}$ are $i$-periodic functions (with the additional interchange of some components in the nonsymmetric case) on a Riemann sheet defined with short cuts, which can be rewritten as
\beqa
\omega_{ab}(u+i) = \omega_{ \bar{a} \bar{b} }(u) ,
\eeqa
where $a = \bar{a}$ for $a =1, \dots, 4$ and $\bar{5} = 6$, $\bar{6} = 5$. The physical meaning of this second system and its r$\hat{\text{o}}$le in the derivation of the Asymptotic Bethe Ansatz equations will be clarified
in \cite{ABJMLong}.
\subsection{ Identification with $\mathcal{N}=4$ SYM }
An interesting formal identification is possible between (\ref{nonsymConstr}-{\ref{eq:Pmu2}) and the ${\bf P }\mu$-system previously derived for the $\mathcal{N}=4$ SYM spectral
problem~\cite{PmuAdS5,AdS5Long}. This can be found by parametrising the ABJM matrix $\mu_{ab}$ in terms of 8 functions
$\nu_i$, $\bnu_i$, $i = 1, \dots , 4$ as follows:
\begin{widetext}
{
\footnotesize
\beq\label{eq:parametrise}
\mu_{a b}=
\left(
\begin{array}{cccccc}
 0 & \nu_1 \bnu_{1} & \nu_2 \bnu_{2} &  \bnu_2 \nu_{3}-\bnu_1 \nu_{4} & \nu_1 \bnu_2 & \bnu_1 \nu_2 \\
 -\nu_1 \bnu_{1} & 0 & \bnu_2 \nu_3 + \nu_1 \bnu_4 & \nu_3 \bnu_3  & \nu_1 \bnu_3 & \bnu_1 \nu_3 \\
 -\nu_2 \bnu_{2} & - \bnu_2 \nu_3 - \nu_1 \bnu_4 & 0 & \nu_4 \bnu_4  & -\bnu_2 \nu_4 & -\nu_2 \bnu_4 \\
  \bnu_1 \nu_{4} -\bnu_2 \nu_{3}  & -\nu_3 \bnu_3 & -\nu_4 \bnu_4 & 0  & -\bnu_3 \nu_4 & -\nu_3 \bnu_4\\
 -\nu_1 \bnu_2 &  -\nu_1 \bnu_3 &  \bnu_2 \nu_4 &  \bnu_3 \nu_4 & 0 & \bnu_2 \nu_3 -\nu_2 \bnu_3 \\
    -\bnu_1 \nu_2 & -\bnu_1 \nu_3  &  \nu_2 \bnu_4 &  \nu_3 \bnu_4 & \nu_2 \bnu_3 - \bnu_2 \nu_3 & 0
\end{array}
\right) ,
\eeq
}
\end{widetext}
with the additional requirement that
\beqa\label{eq:constraint2}
\nu_1 \bnu_4 - \bnu_1 \nu_4 = \nu_2 \bnu_3 - \bnu_2 \nu_3 .
\eeqa
By definition, $\nu_i$ and $\bnu_i$ have the same analytic properties as $\mu_{ab}$, namely $\widetilde{\nu}_i(u) = \nu_i(u+i)$.
The parametrisation (\ref{eq:parametrise}-\ref{eq:constraint2}) is introduced in order to resolve the constraint $\mu \chi \mu \chi = 0$. Moreover, as we discuss below, we expect that $\nu_1^{[+1]}$ and $\bnu_1^{[+1]}$ will play the r$\hat{\text{o}}$le of fundamental Q functions
at weak coupling.
Remarkably, it is possible to rewrite equations (\ref{eq:Pmu1}-\ref{eq:Pmu2}) eliminating $\mu_{ab}$ completely. In fact, one can check that all conditions (\ref{eq:Pmu2})
 are satisfied provided $\nu_i$ and $\bnu_i$ transform in the following simple way under analytic continuation:
\beqa\label{eq:tnu}
\widetilde{\nu_i } = U_{i}^{\; j } \bnu_j , \hspace{2cm}
\widetilde{\bnu_i } = \bar{ U }_{i}^{\; j } \nu_j ,
\eeqa
where
{ \footnotesize
\beqa
{U_{a}}^{b} =
\left( \begin{array}{cccc} { \bf P }_5 & - { \bf P }_2 & { \bf P }_1 & 0 \\ { \bf P }_3 & -{ \bf P }_6 & 0 & { \bf P }_1 \\ { \bf P }_4 & 0 & -{ \bf P }_6 & { \bf P }_2 \\ 0 & { \bf P }_4 & - { \bf P }_3 & { \bf P }_5 \end{array} \right), \;\;
{\bar{U}_{a}}^{ \; b } =
\left( \begin{array}{cccc} \bP_6 & - \bP_2 & \bP_1 & 0 \\ \bP_3 & - \bP_5 & 0 & \bP_1 \\ \bP_4 & 0 & - \bP_5 & \bP_2 \\ 0 & \bP_4 & - \bP_3 & \bP_6 \end{array} \right). \nn
\eeqa
}
Finally, the discontinuity relations for $\bP_i$ can be rewritten as
\beqa\label{eq:tP2}
{ \tilde \bP}_1 - \bP_1 &=& \nu_2 { \widetilde \nu }_1 - \nu_1 { \widetilde \nu }_2  , \;\;
{ \tilde \bP}_2 - \bP_2 = \nu_3 { \widetilde \nu }_1 - \nu_1 { \widetilde \nu }_3 , \nn\\
{ \tilde \bP}_3 - \bP_3 &=& \nu_4 { \widetilde \nu }_2 - \nu_2 { \widetilde \nu }_4 , \;\;
{ \tilde \bP}_4 - \bP_4 = \nu_4 { \widetilde \nu }_3 - \nu_3 { \widetilde \nu }_4 , \nn\\
{ \tilde \bP}_5 - \bP_5 &=& \nu_4 { \widetilde \nu }_1 - \nu_1 { \widetilde \nu }_4 , \;\;
{ \tilde \bP}_6 - \bP_6 = \nu_3 { \widetilde \nu }_2 - \nu_2 { \widetilde \nu }_3  .\hspace{1cm}
\eeqa
To present the identification with $\mathcal{N}=4$ SYM, for simplicity let us restrict to the symmetric sector, by taking $\nu_i = \bnu_i$ and $\bP_5 = \bP_6$. Defining $
\bP_i^{ \mathcal{N}=4 }
:= \nu_i $ for $i = 1, \dots , 4$ and organising the components $\bP_j$ into a
$4\times 4$ anti-symmetric matrix $
\mu_{ab}^{ \tiny \mathcal{N}=4 }
$ as shown in Table \ref{tab:table}, one can see that, on the algebraic level, equations (\ref{eq:tnu}-\ref{eq:tP2})
are \emph{identical} to the Quantum Spectral Curve equations for the left/right-symmetric sector of $\mathcal{N}=4$ SYM~\cite{PmuAdS5}!
Even the constraints perfectly match: in fact notice that (\ref{eq:constr2}) translates into the constraint of \cite{PmuAdS5}:
\beqa
\hspace{-0.8cm}({\mu_{23}}^{{ \tiny \mathcal{N}=4 }})^2 = 1 + {\mu_{13}}^{{ \tiny \mathcal{N}=4 }}{\mu_{24}}^{{ \tiny \mathcal{N}=4 }} - {\mu_{12}}^{{ \tiny \mathcal{N}=4 }}{\mu_{34}}^{{ \tiny \mathcal{N}=4 }} .
\eeqa
Even in the non parity-invariant case, we found an identification with the $\bP\mu$-system for the most general nonsymmetric sector of $\mathcal{N}=4$ SYM, described in \cite{AdS5Long}.
Fascinatingly, the two theories differ only in the analytic properties. As one can see from Table \ref{tab:table}, one could transform the ABJM model into $\mathcal{N}=4$ SYM simply by exchanging the two types of cut structures presented in Figure \ref{fig:cuts}, so that $i$-periodic functions $\leftrightarrow$ functions with a single cut.
\begin{table}
\begin{tabular}{|c|c|}
\hline
 $\mathcal{N}=4$ SYM & ABJM \\
\hline
$
\mu_{ij}
$, \;\;\;{ \scriptsize $i, j = 1, \dots , 4$ } & $\left( \begin{array}{cccc} 0 & -{ \bf P}_1 & - { \bf P }_2 & -{ \bf P}_0 \\
{ \bf P}_1 & 0 & -{ \bf P}_0 & -{ \bf P}_3
\\ { \bf P}_2 & { \bf P}_0 & 0 & -{ \bf P}_4
\\ { \bf P}_0 & { \bf P}_3 & { \bf P}_4 & 0
\end{array} \right)$ \\
\hline
$
\bP_i
$ , \;\;\;{ \scriptsize $i = 1, \dots , 4$  } &  $\nu_i$ \\
\hline
\end{tabular}
\caption{ The single-cut $\leftrightarrow$ periodic mapping between ABJM and $\mathcal{N}=4$ SYM (symmetric case), where we have denoted $\bP_5 = \bP_6 = \bP_0$.}
\label{tab:table}
\end{table}
\section{ Description of the spectrum }
In this Section we provide the information needed to study the subsector of the ABJM model 
which includes the states dual to a folded spinning string with angular momenta $L$ in $CP^3$ and $S$ in $AdS_4$. The subsector is completely characterised by the pair of integers ($L$,$S$) and by the conformal dimension $\Delta$. 
In the $\bP\mu$-system, these quantum numbers are encoded in the asymptotics.
In particular, as observed in \cite{SolvingYsystem} in the $\mathcal{N}=4$ case, $\Delta$ appears in the large-$u$ behaviour
of the product of Y functions $Y_{1,1} Y_{2,2} $:
\beqa
\ln Y_{1,1} Y_{2,2}(u) = 2 i \frac{( \Delta - L )}{u} + \text{O}(\frac{1}{u^2}).
\eeqa
This quantity can be computed as
\beqa \ln Y_{1,1} Y_{2,2}(u) = \ln \mu_{12}(u+i) - \ln \mu_{12}(u) \sim i \partial_u \ln \mu_{12}(u), \nn \eeqa
 and this implies that 
\beqa\label{eq:asymu}
\nu_1(u) = \sqrt{\mu_{12}(u) } \simeq  u^{\Delta - L }.
\eeqa
The asymptotics of $\bf P$ functions is related to the $CP^3$ momentum $L$ as
\beq\label{eq:asyP}
\bP_a(u) \sim ( A_1 u^{-L} , A_2 u^{-L-1} , A_3 u^{+L+1} , A_4 u^{+L}) ,
\eeq
with $\bP_5 = \bP_6 = \sqrt{ 1 + \bP_2 \bP_3 - \bP_1 \bP_4 }$.
To complete the description of the state, we need the following relations between the coefficients $A_i$:
{
\beqa\label{eq:AA}
A_1A_4&=& -\frac{ ((\Delta -S+1)^2 -L^2)((\Delta + S )^2 -L^2 )}{L^2 (2
   L+1)} , \\
A_2 A_3&=&-\frac{((\Delta-S + 1 )^2 - (L+1)^2) ( (\Delta + S )^2 -(L+1)^2  )}{(L+1)^2
   (2 L+1)} . \nn
\eeqa
Equations (\ref{eq:AA}) can be derived as discussed in  \cite{PmuAdS5, AdS5Long, ABJMLong}. It is interesting that, as remarked in \cite{QSCAtWork}, the quantisation of $S$ appears naturally through the nonlinearity of the $\bP\mu$-system. The identifications above involve some guesswork, but they can be checked 
by recovering the correct weak coupling result, as shown in the next section. In principle, equations (\ref{eq:asymu}-\ref{eq:AA}) are the \emph{only} physical input needed for the computation of $\Delta$ at any value of $h$.
}
\subsection{A weak coupling test}
As a test of our results, let us show that they reproduce the $2$-loop Baxter equation. At leading order at weak coupling, we expect that
\beqa
\Delta = L + S +\text{O}(h^2),
\eeqa
and we see from (\ref{eq:AA}) that $A_2 A_3 = \text{O}(h^2)$.
Therefore we assume that $ \bP_2 \to 0 $, and we see that as a consequence the equations for $\nu_1$ and $\nu_3$ decouple:
\beq\la{nutred}
\left( \begin{array}{c} { { \tilde \nu } }_1 \\ { { \tilde \nu } }_3  \end{array} \right) =\left( \begin{array}{c} {  \nu }^{[+2]}_1 \\ { \nu}^{[+2]}_3  \end{array} \right) =
 \left( \begin{array}{cccc} \bP_0 & \bP_1  \\
\bP_4 &  -\bP_0  \end{array} \right) \left( \begin{array}{c} \nu_1
\\ \nu_3  \end{array} \right).
\eeq
Making the identification $\nu_1^{[+1]} = Q$, the system (\ref{nutred}) implies the Baxter equation:
\beqa
\left(  \frac{\bP_0^{[+1]}  }{\bP_1^{[+1]}  } - \frac{\bP_0^{[-1]}  }{\bP_1^{[-1]}  }  \right) Q = \frac{  Q^{[-2]}  }{\bP_1^{[-1]}} -\frac{Q^{[+2]}}{\bP_1^{[+1]}}.
\eeqa
Generalising the argument of~\cite{PmuAdS5},
one can go further and reproduce the expected $2$-loop result~\cite{MinahanZarembo}:
\beqa
\Delta &=& L + S + 2 i h^2 \partial_u \log \left.\frac{ Q^{[+1]} }{ Q^{[-1]} }\right|_{u=0}
+ \text{O}(h^4) .
\eeqa

\section{Conclusions}
In this paper we have recast the spectral problem for the ABJM model as a finite system of coupled Riemann-Hilbert equations:
 the $\bP \mu$-system. The similarity with the $\mathcal{N}=4$ SYM case suggests that an analogous formulation should exist also
 for the, still partly mysterious, integrable models related to $AdS_3/CFT_2$. Studying other examples would probably help to understand the hidden algebraic structures underlying these systems. It would be particularly interesting to investigate how the analytic properties of the $\bP\mu$-system are modified under the $q$-deformation discussed in \cite{QDeformed}. This may help to clarify the physical meaning of the formal map between the QSC equations for $\mathcal{N}=4$ SYM and ABJM presented in this letter.

Let us summarise some of the potential applications to ABJM.
Adapting the methods of \cite{9loops,PmuAdS5,QSCAtWork}, our results should allow to study the weak and strong coupling
expansions, and non-perturbative near-BPS regimes such as the small-spin limit described by the slope function~\cite{ABJMSlope}.
An interesting open problem would be to find numerical solution methods valid at generic values of the coupling. We believe that our equations can also be applied to study the spectrum of cusped Wilson lines.

Finally, one can hope that studying the $\bP\mu$-system in the ABJM context would reveal some structures which are harder to see in the case of ${\cal N}=4$ SYM and help to clarify
 the nature and the r$\hat{\text{o}}$le of this intriguing mathematical object
 both in the AdS/CFT correspondence and in the general theory of integrable models. 
Hopefully, this can also
 teach us something new about
non-perturbative gauge theories and AdS/CFT.
  \begin{acknowledgments}
\section*{Acknowledgements}
\label{sec:acknowledgments}
We would like to thank L. Bianchi, D. Bombardelli, F. Levkovich-Maslyuk, S. Negro, G. Sizov, S. Valatka for many useful discussions, and we especially thank V. Kazakov, S. Leurent and D. Volin for sharing with us the draft of~\cite{AdS5Long}. 
This project was partially supported by INFN grants IS FTECP, IS GAST,
the UniTo-SanPaolo research grant Nr TO-Call3-2012-0088, the ESF Network
09-RNP-092 (PESC) and and MPNS COST Action MP1210.
The research of N.G. leading to these results has received funding from the
People Programme (Marie Curie Actions) of the European Union's Seventh
Framework Programme FP7/2007-2013/ under REA Grant Agreement No 317089.
N.G. wishes to thank the STFC for partial support from the consolidated grant ST/J002798/1.
\end{acknowledgments}

\ifbibtexelse{
\bibliography{biblio.bib}
}

\end{document}